\documentclass[10pt]{bmc_article}   

\usepackage{amsmath,amsfonts,amssymb}
\usepackage[pdftex]{color,graphicx,hyperref}
\usepackage{subfigure}
\usepackage{graphicx}
\usepackage{multirow}
\usepackage{array}
\usepackage{colortbl}

\usepackage{ifthen}
\usepackage{cite}
\usepackage{url}
\usepackage{multicol}

\usepackage[utf8]{inputenc}

\newboolean{publ}

\newenvironment{bmcformat}{\baselineskip20pt\sloppy\setboolean{publ}{false}}{\baselineskip20pt\sloppy}

\begin{document}

\begin{bmcformat}

\title{Effects of time window size and placement on the structure of aggregated networks}

\author{Gautier Krings\correspondingauthor$^1$%
         \email{Gautier Krings\correspondingauthor - gautier.krings@uclouvain.be}
       \and
       M\'arton Karsai$^2$%
       \and
       Sebastian Bernhardsson$^3$%
       \and
       Vincent D. Blondel$^1$ %
       \email{Vincent D. Blondel - vincent.blondel@uclouvain.be}
       and 
       Jari Saram\"a{}ki\correspondingauthor$^2$%
       \email{Jari Saram\"aki\correspondingauthor - jari.saramaki@aalto.fi}
      }
\address{%
    \iid(1)ICTEAM Institute, Universit\'e{} catholique de Louvain, Avenue Georges Lema\^i{}tre, 4, 1348 Louvain-la-Neuve, Belgium\\
    \iid(2)Department of Biomedical Engineering and Computational Science, Aalto University School of Science, P.O. Box 12200, FI-00076, Finland\\
    \iid(3) Swedish Defence Research Agency, SE - 147 25 Tumba, Sweden
}%

\maketitle
\begin{abstract}
Complex networks are often constructed by aggregating empirical data over time, such that a link represents the existence of interactions between the endpoint nodes and the link weight represents the intensity of such interactions within the aggregation time window.
The resulting networks are then often considered static. More often than not, the aggregation time window is dictated by the availability of data, and the effects of its length on the resulting networks are rarely considered. Here, we address this question by studying the structural features of networks emerging from aggregating empirical data over different time intervals, focussing on networks derived from time-stamped, anonymized mobile telephone call records. Our results show that short aggregation intervals yield networks where strong links associated with dense clusters dominate; the seeds of such clusters or communities become already visible for intervals of around one week. The degree and weight distributions are seen to become stationary around a few days and a few weeks, respectively. An aggregation interval of around 30 days results in the stablest similar networks when consecutive windows are compared. For longer intervals, the effects of weak or random links become increasingly stronger, and the average degree of the network keeps growing even for intervals up to 180 days. The placement of the time window is also seen to affect the outcome: for short windows, different behavioural patterns play a role during weekends and weekdays, and for longer windows it is seen that networks aggregated during holiday periods are significantly different. 
\end{abstract}

\ifthenelse{\boolean{publ}}{\begin{multicols}{2}}{}

\section*{Introduction}

Complex networks have become a standard tool for representing the interaction structure of complex systems~\cite{Newman06a,
NewmanBook11}. The strength of the network approach comes from its ability to cast the essential features of increasingly complex systems into a manageable form -- in the simplest representation, interacting elements are mapped to nodes that are connected by links if they are known to interact. While this coarse-grained view has given a lot of insight into the key characteristics of such systems, it is evident that it entails several approximations and underlying assumptions. The first is the criterion for the existence of links -- if the interactions are not binary (on/off) by nature, when is an interaction strong enough to be represented as a link? A common way of taking such strengths into account is to assign weights to the links of the network~\cite{barrat2004acw}.  The second approximation is related to the time domain. Standard network theory deals with networks that are either static or only slowly changing in time. However, in reality, there are typically dynamical changes in the network structure on multiple time scales. Consequently, representing an empirical system as a static network involves aggregating or integrating over the network dynamics over some time interval. In addition, in many cases, the interactions of the system are not continuously active. 
While the microdynamics of link activations may be taken into account with the temporal network framework~\cite{PetterJari}, for the aggregated network approach, the interaction frequencies are often used to define the edge weights. 
It is evident that when aggregating interactions over time, the choice of the aggregation window and its length have consequences on the characteristics of the resulting networks~\cite{Holme2003}. However, this issue has often been neglected in the literature; often, the aggregation interval has been dictated by the availability of data,  while it would be beneficial to ensure that the network properties that one is interested in are captured by the aggregated networks. 

In this paper, we address this question by monitoring and analyzing the features of network structure emerging from aggregation over different time intervals for an empirical data set human communication.
We present a detailed study of the effects of the aggregation window on the structural features human communication networks that are known to display dynamics on multiple overlapping time scales. The data comes in the form of a time-stamped sequence of mobile telephone calls between anonymized customers of a Belgian mobile operator for a period of 6 months. This sequence is then aggregated over time to form links between customers, and key features of the resulting networks are studied. 

There is an increasing number of studies of human social networks derived from telecommunication records. However, the networks analyzed in the literature have been constructed using very different time windows -- a day~\cite{aiello}, a week~\cite{nanavati}, one month~\cite{seshadri}, and several months (e.g.~\cite{onnela2007sat,lambiotte2008gdm}) -- and therefore it is crucial to understand what features of the underlying system are captured by different aggregation intervals. 
For such social communication networks, there are several mechanisms that are expected to affect the resulting network structure.
First, the distribution of link weights, \emph{i.e.}~call frequencies, is broad~\cite{onnela2007sat,onnela2007als}. Thus there are high-weight links that should on average be observed earlier on in the aggregation process, and many links of low weight that take a long time to be observed. Second, link weights are correlated with network topology, such that high-weight links are associated with denser network neighbourhoods~\cite{onnela2007sat}. Third, for links of any weight, it is known that the distributions of inter-call times are also broad, \emph{i.e.}~call sequences are bursty~\cite{Candia2008,Karsai11}, giving rise to longer-than-Poissonian waiting times between calls. Fourth, there are circadian patterns~\cite{JoArxiv2010}, where the overall level of call activity varies by hour, as well as weekly patterns where call behavior depends on the day of the week. Fifth, there are changes in the network itself too -- relationships grow and wane in strength, new links appear, and old ones are terminated. The aggregated network structure then reflects the joint effect of the above mechanisms that are associated with different time scales. Thus, one cannot expect that there is a proper aggregation interval that represents the true network; rather, different structural features emerge with different aggregation times. In order to understand what the network structure represents, it is important to understand this process.

This paper is structured as follows: first, we discuss the structural and temporal inhomogeneities that are expected to affect the features of aggregated networks. Then, 
we characterize the dependence of fundamental scalar measures of network structure on the aggregation interval, and address the properties of links added at different times during the aggregation procedure. We find that clustering of the network peaks at 9 days, as the strongest links associated with dense clusters are observed early on in the process. Another time scale is related to the stability of the aggregated networks -- networks aggregated for around 30 days display the largest similarity between consecutive windows. Moving from scalar measures to distributions, we find that the degree and weight distributions become surprisingly stationary in 1-2 weeks of aggregation time. Finally, we investigate in detail the effects of different aggregation window placements, and show that the underlying behavioural patterns affect the aggregated networks: on short time scales, weekends differ from weekdays, and on longer scales, holiday periods give rise to anomalies in the aggregated network structure.

\subsection*{Data}

Our data consist of the anonymized mobile telephone call records of the customers of a Belgian mobile operator from October 1, 2006 to March 31, 2007. Each customer is uniquely identified, and each call is associated with a time stamp and a duration.
This data set has already been studied from a static perspective in several papers \cite{lambiotte2008gdm,krings2009urban,blondel2010regions}. As our focus is on link dynamics, we filter out all customers who have modified their subscription plan during the data collection period. This removes new customers, and customers who have cancelled their subscription during the period.
This leaves us with a network of 2.1 millions customers, making over 170 millions calls during the collection period.

For reference, we also construct two randomized ensembles, based on two randomization techniques of the time stamps. For both cases, the resulting randomized reference sequences contain the same number of calls between the same individuals as the original data. In the first ensemble, the time stamps of all calls are generated uniformly at random over the complete time range, in order to remove the system-level call frequency pattern (daily and weekly pattern). In the second ensemble, the time stamps of all calls are randomly reshuffled, which retains the daily and weekly patterns, but removes other temporal correlations between the timings of calls of links. When aggregating over the entire observation period, the call sequences from both reference models produce networks that are equal to the network from aggregating the original data. In the remaining, we will refer to these references as respectively the ``uniform'' and the ``shuffled'' references.

\section*{Network growth}


\begin{figure}[t]  \begin{center}
  \includegraphics[width=1.0\linewidth]{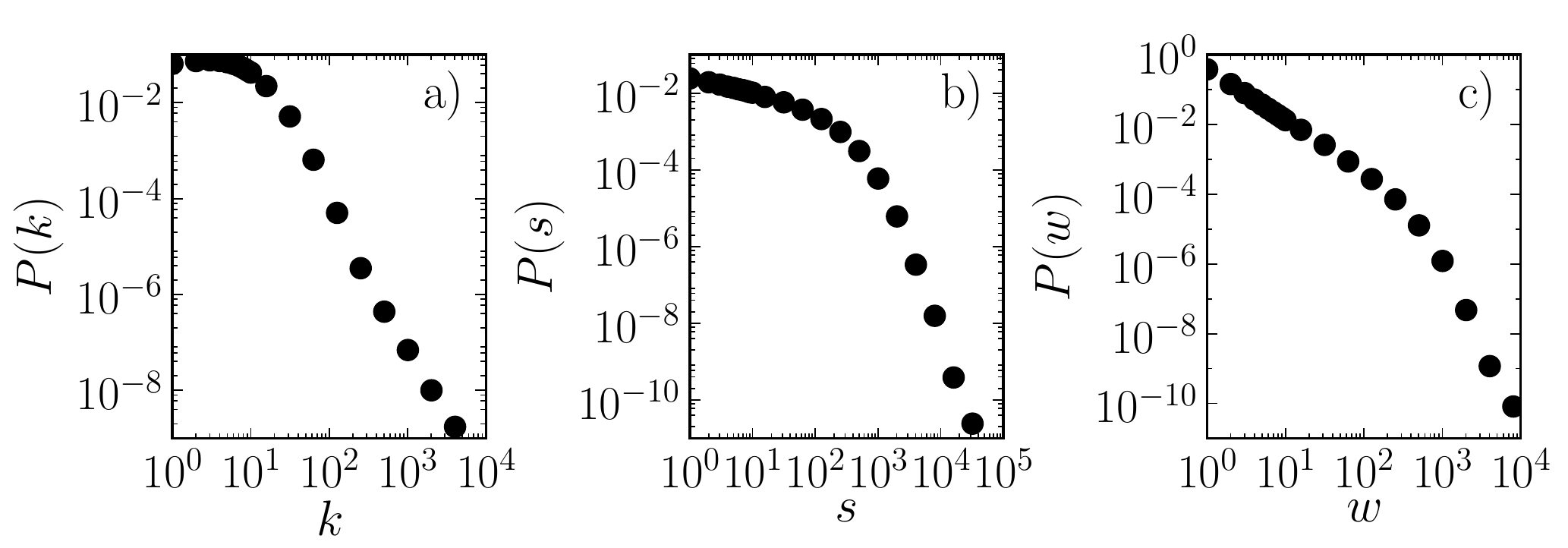} 

 \end{center}
 \caption{The degree (a), strength (b) and weight (c) distributions of the aggregated network when the aggregation period covers 
the whole 6 months of data.}
 \label{fig:distributions}
\end{figure}

\subsection*{Structural and temporal inhomogeneities}

We begin our investigations by addressing the inhomogeneities that are expected to play an important role in the evolution of the structural properties of networks aggregated over growing time intervals. The fundamental structural inhomogeneities are reflected in the standard statistical distributions for the call network, aggregated over the entire 6-month period of observation. In the aggregated network $G(t)$, a link is established between nodes $i$ and $j$ if a call is observed between them at any point during the aggregation interval $(0,t)$; the weight $w_{ij}$ of the link is defined as the total number of calls between $i$ and $j$ within the interval. The strength $s_i$ of node $i$ is then defined~\cite{barrat2004acw} as the total number of calls where $i$ participates, and the degree $k_i$ as usual as the number of links that node $i$ has. As expected on the basis of earlier results~\cite{onnela2007sat,onnela2007als,lambiotte2008gdm}, the probability density distributions of degree, strength, and link weights are all broad (see Figure 1). Thus there is a large number of nodes that make only infrequent calls, and a large number of links that carry only a few calls. When aggregating the network over shorter time intervals, one thus expects to first discover the high-strength nodes and high-weight links that are associated with the tails of these distributions. 

In the time domain, the two main inhomogeneities are related to burstiness of calls forming the links, and the overall circadian pattern of the system-wide call frequency.
Burstiness of the calls is reflected in the probability distribution $P(\tau)$ of the times $\tau$ between consecutive calls on individual edges. In Fig. 2 a), it is seen that in line with earlier observations \cite{Candia2008,Karsai11,Miritello2011}, the distribution $P(\tau)$ in our empirical data has a broader-than-Poissonian tail, a signature of burstiness. Such an inter-call time
distribution gives rise to longer waiting times than expected if the calls were placed uniformly in time. Because of this, we expect to see slower network growth than for the uniform case. 
Further, as seen in Fig. 2, the network-level call frequency clearly displays the usual daily and weekly pattern~\cite{Karsai11,JoArxiv2010}, where the frequency shows two daily peaks followed by a decrease during nights. In addition, weekend activity is lower, especially for Sundays. 

\begin{figure}[t]  \begin{center}
\includegraphics[width=1.0\linewidth]{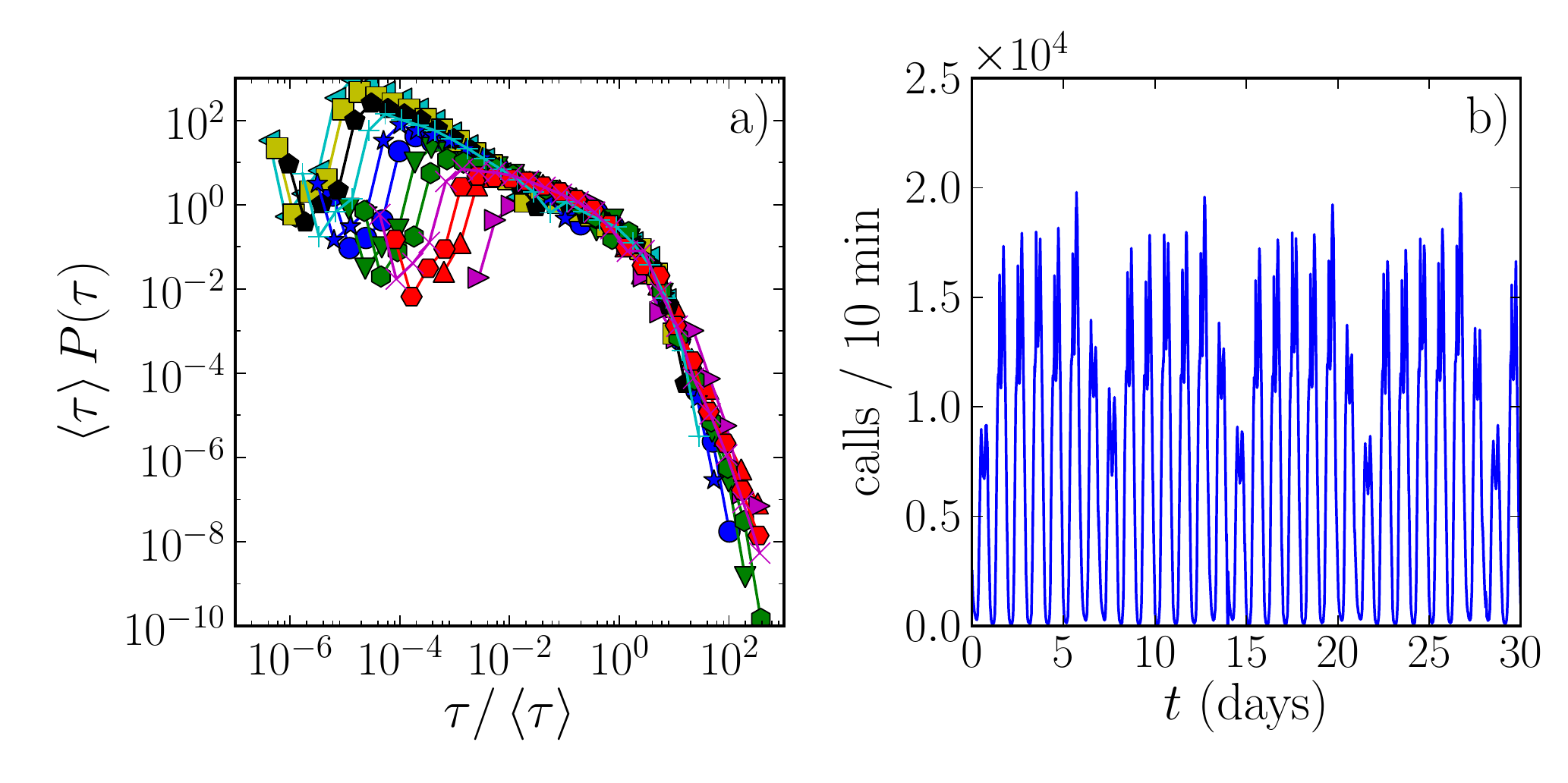}
 \end{center}
 \caption{Temporal inhomogeneities affecting network growth. a) The scaled inter-event time distribution of links $P(\tau)$, displaying a broader-than-Poissonian tail. For the plot, the edges have been divided into bins of different numbers of calls. The inter-event time distributions are then scaled by the average inter-event time $\left<\tau\right>$ of each bin, following~\cite{Candia2008}. The non-scaling regime for low $\tau$ can be attributed to correlated calls, where an incoming
 call triggers an outgoing call within a short time period.
b) Total call rate in the network as a function of time, for the first 30 days, displaying the circadian and weekly patterns. Deep drops can be observed at night, while there are daily peaks, the highest of which appear on Friday evenings. Overall call activity is lower in the weekends, especially on Sundays.}
 \label{fig:eventrate}
\end{figure}

\subsection*{Evolution of network structure}

All of the above features are expected to have an effect on the properties of networks aggregated over growing time intervals. Let us first monitor the growth of the aggregated network in terms of the numbers of nodes and links and the average degree, when the network $G(t)$ is aggregated up to a time $t$. As seen in Fig. 3 a), the number of observed nodes $N(t)$ displays a rapid increase in the beginning of the aggregation process, such that the aggregated network contains 90$\%$ of the nodes after $t\sim 30$ days. This rapid increase is  followed by slower growth as nodes with low call activity are gradually observed to make calls, joining them to the aggregated network. When compared to the uniform reference, where the time stamps of all calls are drawn uniformly at random from the entire 6-month interval, it is seen that the growth of $N(t)$ is slightly slower; however, for longer aggregation times, the difference can be considered negligible and thus the time-domain heterogeneities have a visible effect only for short time windows. For short time windows, in addition to the slowing-down effect of burstiness, the daily pattern is seen to give rise to a stepped shape of the $N(t)$ curve (see the inset of Fig. 3 a).

\begin{figure}[t]  \begin{center}
  \includegraphics[width=0.5\linewidth]{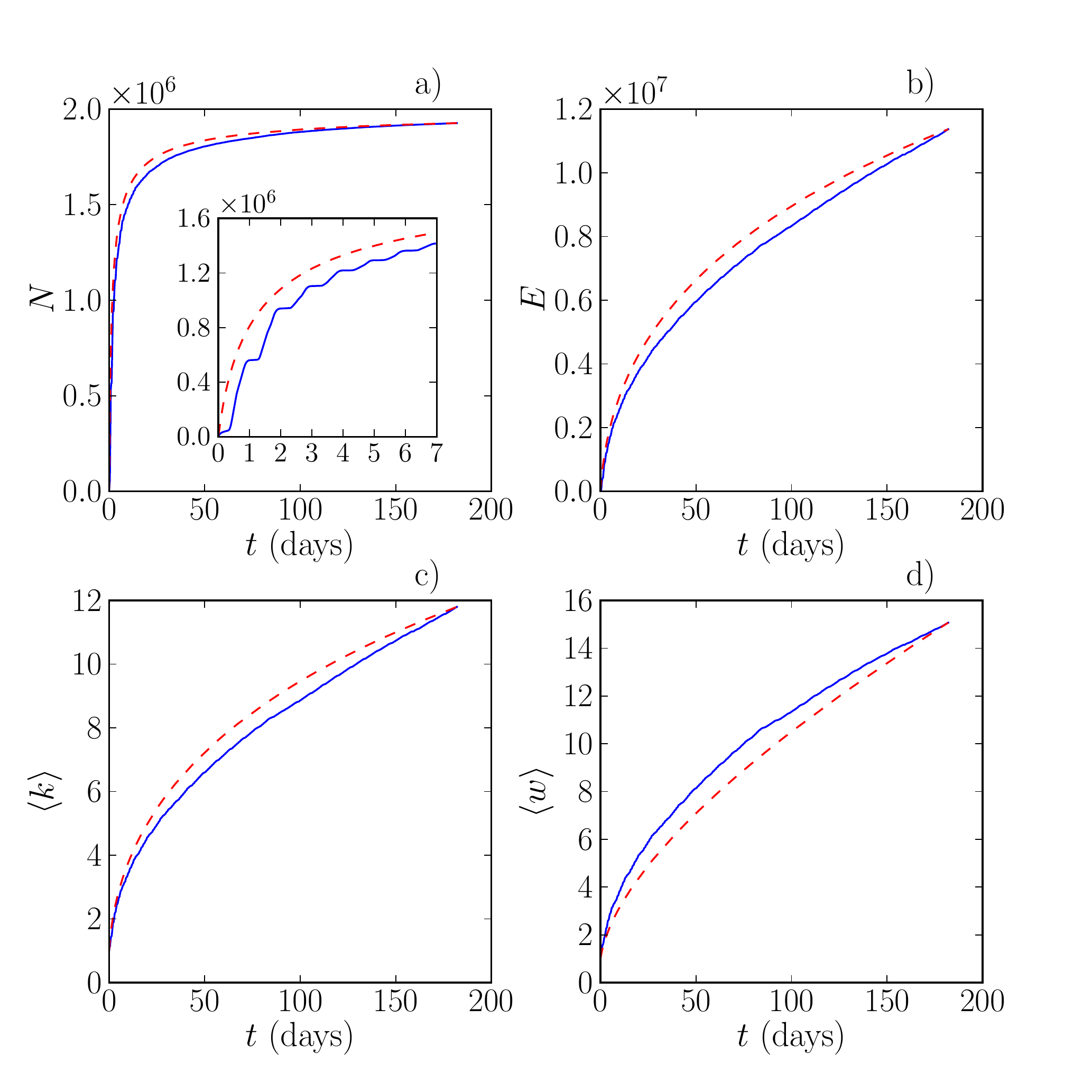}
 \end{center}
 \caption{Total numbers of (a) nodes and (b) edges and (c) the average degree and (d) the average edge weight in the aggregated network as a function of aggregation time. The solid (blue) line denotes original
 empirical data, while the dashed (red) line denotes the uniform reference. The inset in panel (a) displays the number of nodes for the first 7 days.}
 \label{fig:basics}
\end{figure}

\begin{figure}[t]
\begin{center}
\includegraphics[width=0.7\linewidth]{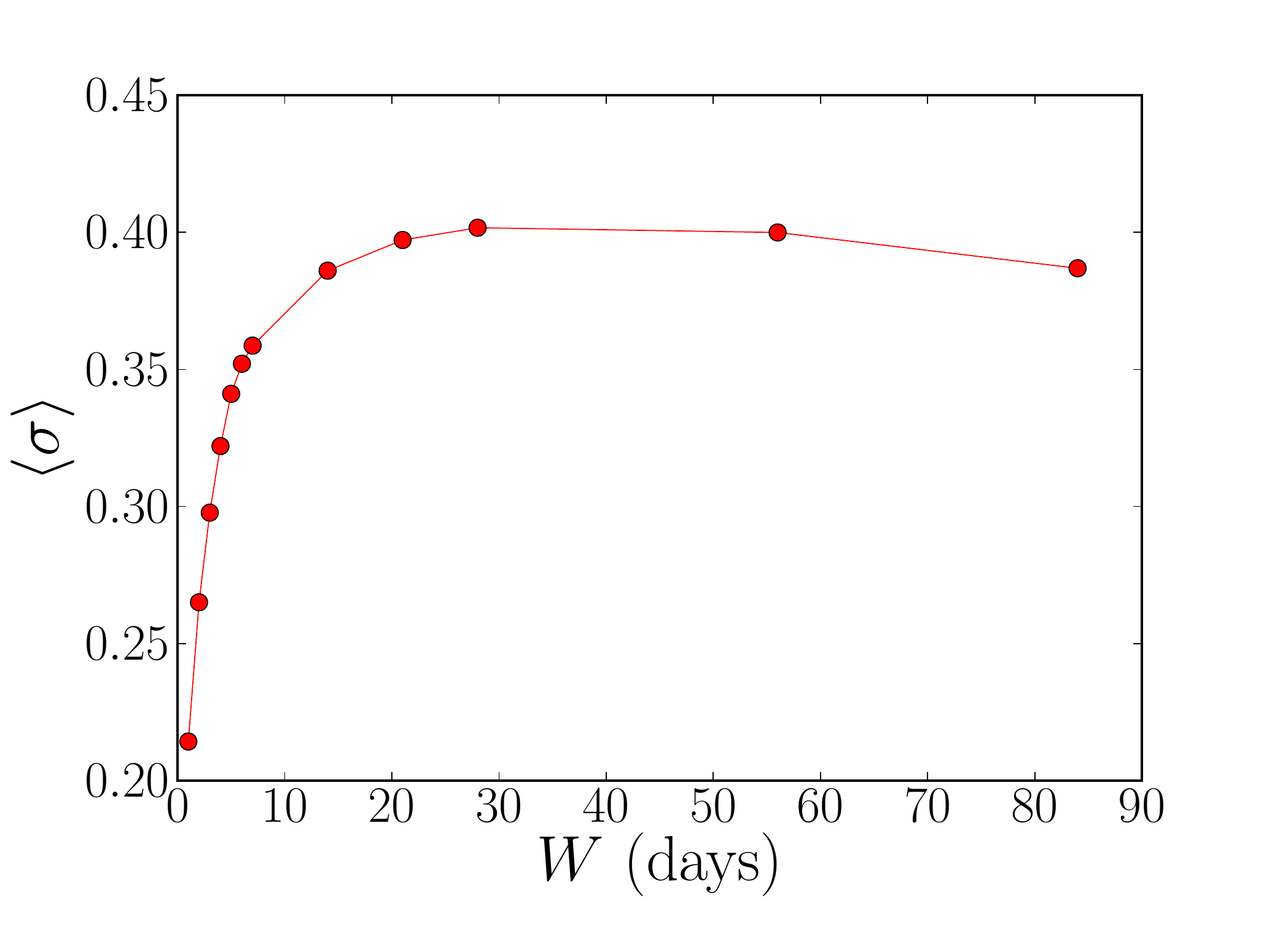}
\caption{The average fraction of links $f$ common to consecutive aggregation windows of duration $W$.}\label{fig:similarity}
\end{center}
\end{figure}

In contrast, the growth in the number of edges $E(t)$ is much more gradual, as seen in Fig. 3 b). Here, an aggregation time of $t\sim 149$ days is required for catching 90$\%$ of the edges of the final 6-month aggregated network. In addition, unlike for the number of nodes, for long aggregation times, the number of edges keeps on growing steadily and no saturation in growth is observed. This is also reflected in the growth of the average degree (Fig. 3 c). Hence, even though the number of nodes becomes fairly stable in  an aggregation period of 6 months, one cannot claim to have captured all the edges of the underlying network, and for longer windows, the average degree would still increase. This reflects the joint effect of several factors: first, as the edge weight distribution is broad, there are large numbers of edges with very low call frequencies, and observing those evidently takes a long time; there may be many edges where calls take place less frequently than once in six months. In addition, the ubiquitous burstiness that results in longer waiting times between calls slows down the growth in the number of links especially for the low-weight links -- this effect is visible in Fig. 3 b), although it is not very strong. Second, for such long observation periods, one can argue that the changes in the network structure should already have a visible effect: new social ties are formed while older ties wane in strength and may even cease to exist. Third, as the data contains all the calls made by the subscribers, many of the calls may be random in the sense that they do not reflect the structure of the underlying social network -- as there is no background information on the nature of the calls, a random call to one's dentist or a call in response to an advertisement on used car sales are counted as links, just as calls to one's friends or relatives. This third mechanism would naturally result in an ever-growing number of links. The average link weights (Fig. 3 d) must necessarily keep on growing, since all new calls on existing edges are added to their link weight. This growth slows down towards the end of the observation period but does not become as linear-looking as the average degree growth; note that the new links giving rise to growing degrees also affect average weights. Comparison with the uniformly random times reference reveals the effect of burstiness -- weights grow faster in the original data because of burstiness, where rapid sequences of calls following one another quickly increase link weights.

As a result of the interplay of the above mechanisms, the network keeps changing while it is being aggregated, and while some of its links are stable in the sense that they remain active for prolonged periods of time, others exist or can be detected only within limited time periods. Then, one may ask what should the aggregation window 
length be for obtaining representative, "backbone" networks that capture the stablest connections in the system?
One way of obtaining a quantitative estimate of the characteristic time scale of network changes is to compare the similarity of networks aggregated for different periods of time when the observation period is divided into multiple consecutive aggregation windows. We calculate the similarity $\sigma$ of two networks $G_1=(V_1,E_1)$ and $G_2=(V_2,E_2)$ as
\begin{equation} \label{eq:sim}
\sigma(G_1,G_2) = \frac{|E_1 \cap E_2|}{|E_1 \cup E_2|},
\end{equation}
such that $\sigma=1$ if the networks are the same, and $\sigma=0$ if they share no links. Fig. 4 displays the average similarity $\sigma$ of networks in consecutive windows of different durations $W$. When the windows are very short, the networks are very sparse and the number of common links is low. Then, the similarity increases with increasing window duration, reaching a maximum at $\sim 30$ days; subsequently, the similarity begins slowly decreasing as the aggregation process captures more and more of the very weak or random links. 

\begin{figure}[t]
 \begin{center}
  \includegraphics[width=1.0\linewidth]{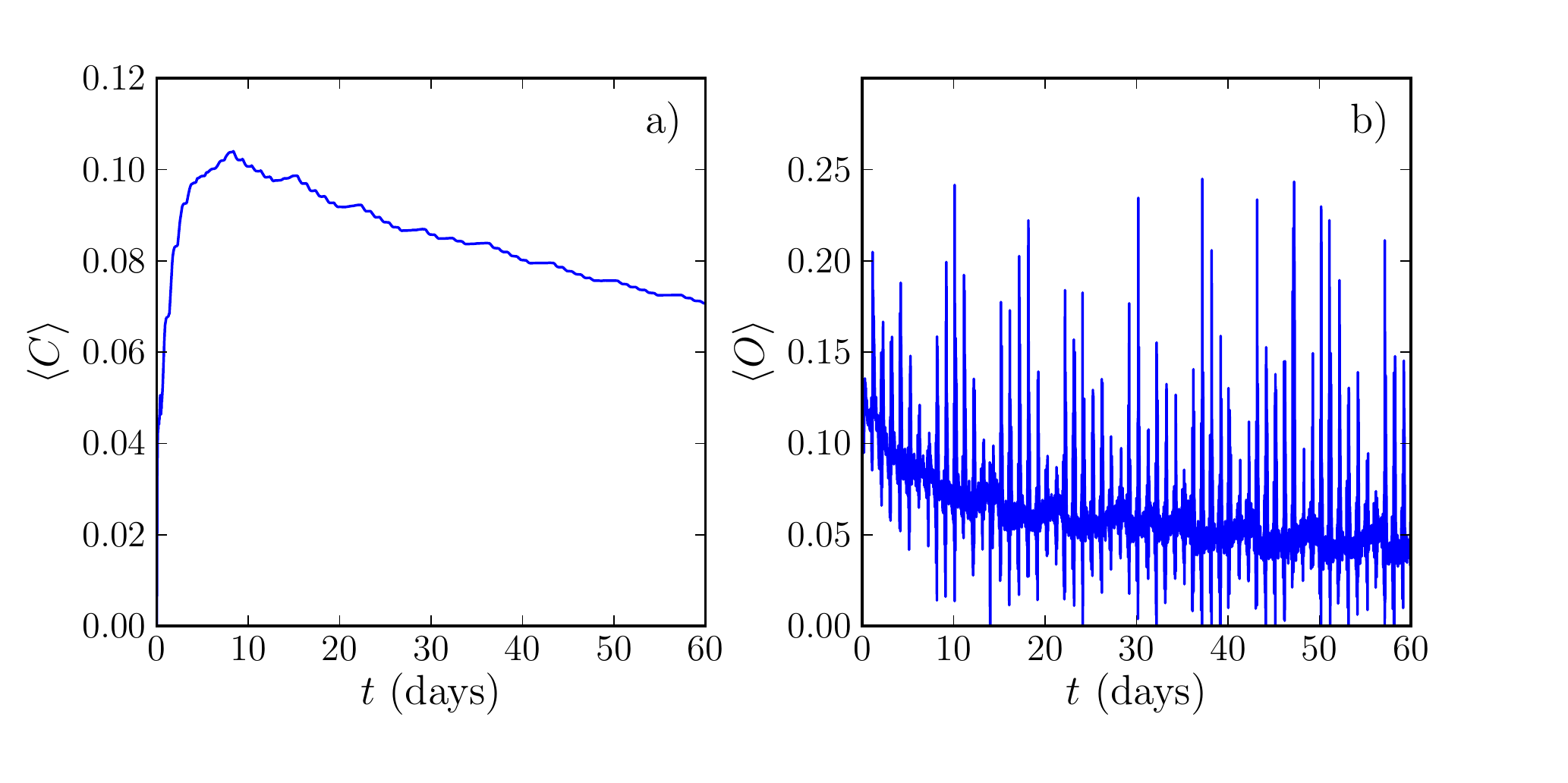}
 \end{center}
 \caption{a) Global clustering coefficient of the network and b) average final overlap of added edges as a function of aggregation time, for the first 2 months. The global clustering coefficient is computed as the number of triangles divided by the number of connected triplets. For the overlap, we calculate the final overlap of edges in the 6-month aggregated network, and average over the final overlap values of newly added links at each time point. }
 \label{fig:clustering}
\end{figure}

\begin{figure}[t]
 \begin{center}
  \includegraphics[width=0.5\linewidth]{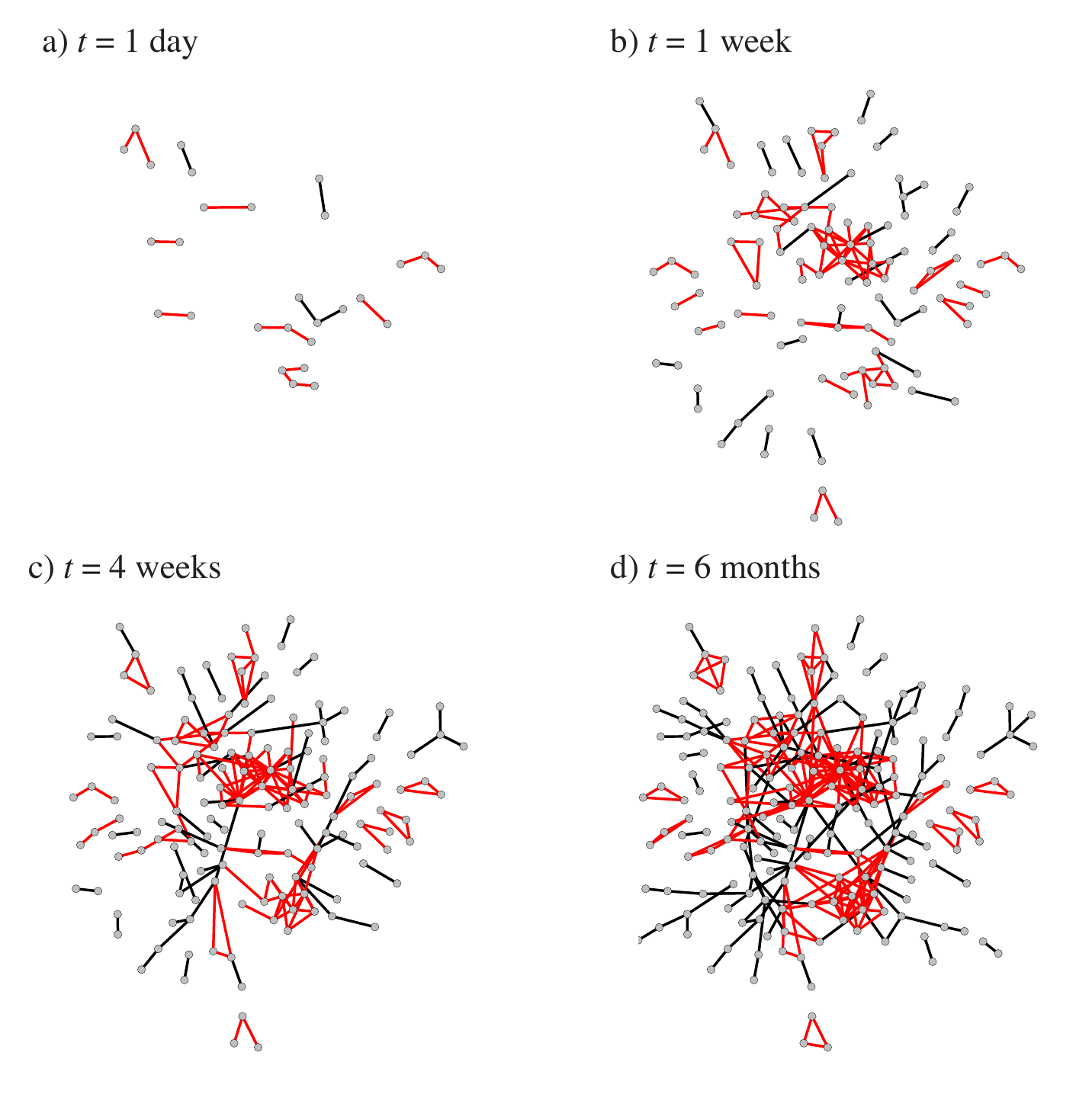}
 \end{center}
 \caption{Series of aggregated networks with a growing aggregation interval. The network aggregated here represents a small subnetwork, obtained by picking individuals from a single postal code. Links that participate in triangles in the final 6-month aggregated network are colored red, while the rest are black. }
 \label{fig:series}
\end{figure}

As the growth of the number of links in Fig. 3 b) does not saturate, it is of importance to understand the characteristics of links that emerge early on in the process.
It is known from previous investigations~\cite{onnela2007sat} that link weights correlate with the network topology such that high-weight links are associated with dense network neighbourhoods, whereas low-weight links connect such neighbourhoods, in line with the Granovetter hypothesis~\cite{GranovetterWeakTies}. This is directly related to the presence of \emph{community structure}~\cite{SantoCommunityReview} in social networks; links within communities are stronger and have higher-than-average weights~\cite{gergelyCommunities2011}. For the network aggregation, this means that clusters and communities containing high-weight links are likely to appear early on in the process. In order to investigate this effect, we measure the evolution of the network-level clustering coefficient $C(t)$, given by $3$ $\times$ the number of triangles divided by the number of connected triplets in the network. As seen in Fig. 5 a), the clustering coefficient does indeed show a rapid increase as a function of the aggregation interval length, and then decreases after a peak at around $t=9$ days. This decrease can be attributed to the weak links observed later in the process: those links contribute less frequently to triangles. Hence, if short aggregation periods of around one week are used, the resulting network structure is dominated by strong links associated with dense clusters.


The fact that the edges observed early on in the aggregation process are related to the community structure is also visible when monitoring the \emph{overlap}~\cite{onnela2007sat} of the added links. The overlap of a link connecting nodes $i$ and $j$ is defined as 
\begin{equation}
O_{ij}=n_{ij}/\left[\left(k_i-1\right)+\left(k_j-1\right)-n_{ij}\right],
\end{equation}
where $n_{ij}$ is the number of common neighbours of $i$ and $j$, and $k_i$ and $k_j$ are their degrees. Thus the overlap measures the fraction of common neighbours out of all neighbours of the two connected nodes. Fig. 5 b) displays the average final 6-month overlap of the added links as a function of aggregation time. Here we have calculated the overlap of each link in the final 6-month aggregated network, and averaged over these values for links that are added to the network at time $t$. It is seen that the links that are added early on in the aggregation process have on average a higher overlap than those added later; the final overlap is a decreasing function. Hence, even when the aggregation times are short, the networks capture features of the community structure of the final aggregated networks. Interestingly, the overlap also shows a strong circadian and weekly pattern -- its highest peaks correspond to the early morning when the overall call rate is very low. Thus, if calls are made during these hours, they are likely to be targeted towards people in the strongest clusters of friends and family. 

In order to illustrate the network growth, we have visualized small subnetworks corresponding to different aggregation times $t$ (Fig. 6).  Here, the subnetwork has been obtained by selecting all individuals whose subscriptions are associated with a certain postal code -- this method of sampling yields 
better results than e.g. snowball sampling. Panels a) to d) of Fig. 6 show the growth of the network, such that edges that participate in triangles in the final 6-month aggregated network are coloured red. For the shortest aggregation periods (panels a and b), most of the added edges are in this set, reflecting the above observations on the early appearance of edges connected to communities and clusters. It should be noted that not all community-internal edges are discovered early on; rather, those links that appear early are associated with communities with a high probability.

%
%

%
\section*{Behaviour of statistical distributions}
\begin{figure}[t]
 \begin{center}
  \includegraphics[width=0.7\columnwidth]{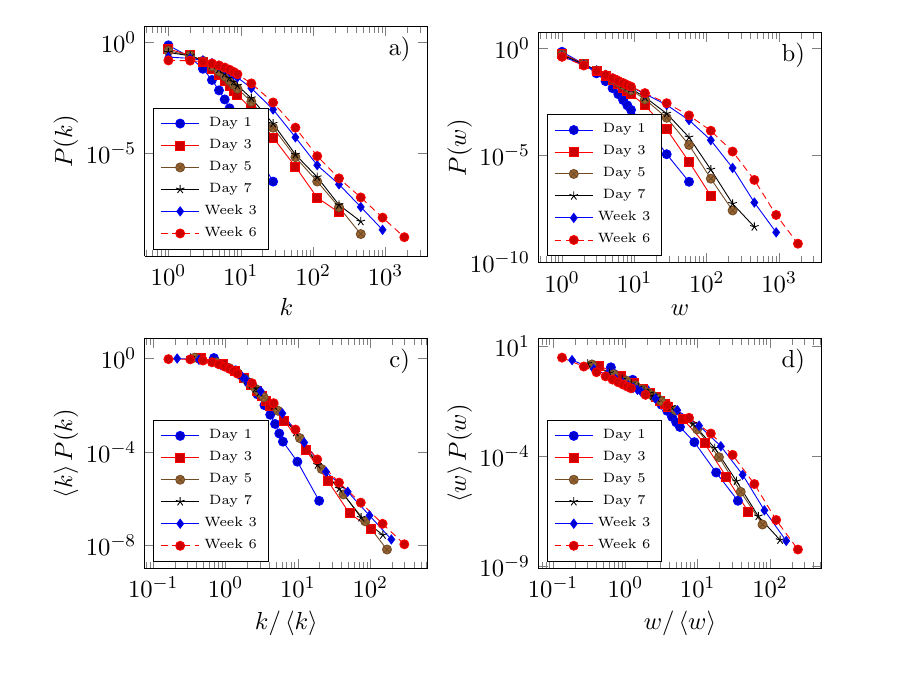}
 \end{center}
 \caption{a) The degree distribution and b) the weight distribution for different aggregation intervals. c) the scaled degree distribution and d) the scaled weight distribution. The distributions are rescaled with respect to their average value.}
 \label{fig:rescaling}
\end{figure}

Above, it was seen that even for fairly long aggregation intervals, the average degree of the aggregated networks still keeps increasing as a function of the interval length. Likewise, because of how the aggregation process is defined, the average weight necessarily keeps on growing as well. Next, we turn to the statistical distributions of these quantities, and ask when they become descriptive of the underlying network. Evidently, as the averages keep increasing with the aggregation interval length, the probability density distributions of degrees and weights also change and shift towards higher degrees/weights. However, for such distributions to be meaningful descriptors of network structure, their underlying forms should for long enough intervals become stationary and depend only on their average values. In order to study the stationarity of the
underlying dynamics one can compare such distributions by rescaling them as 
$P\left(x,\left\langle x\right\rangle\right)\sim\left\langle x\right\rangle P\left(x/\left\langle x\right\rangle\right)$, where $\left\langle x \right\rangle = \sum xP(x)$. Fig. 7 a) and b) display the PDF's for the degrees and weights, respectively, for different aggregation interval lengths. The bottom panels c) and d) display scaled versions of the distribution, rescaled with respect to their average value according to the above.
For degrees, one sees that the rescaled distributions collapse onto a single curve, surprisingly already for aggregation intervals over a few days, while for the weights, the distributions converge slightly slower but eventually also collapse for window sizes of the order of weeks.
\begin{figure}[t]
 \begin{center}
  \includegraphics[width=0.9\columnwidth]{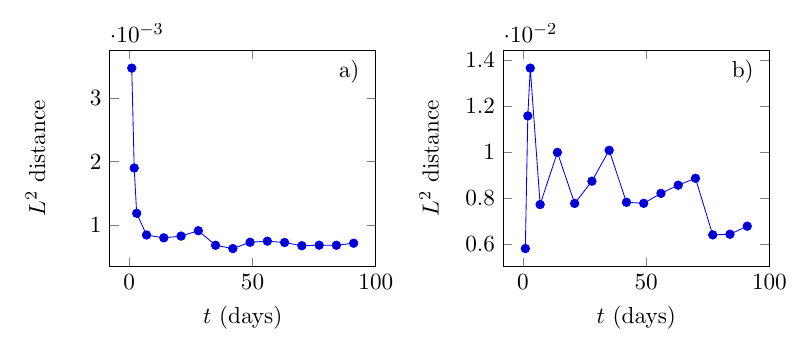}
 \end{center}
 \caption{The $L^2$ distance between the degree (a) and weight (b) distributions of the networks aggregated over a time interval of length $t$ and $2t$. That is, the first point is the distance between the rescaled distributions of the network built with the first day and the network built with the two first days. The second point is the distance between the rescaled distributions of the two first days and the four first days, etc. The last point is the distance between the distributions of the network built with the 91 first days and the network built with the 182 first days.}
 \label{fig:l2}
\end{figure}
The convergence of the rescaled degree distributions onto a single curve suggests that the underlying dynamics that is being aggregated has some stationary properties. A similar observation was made in \cite{Gautreau2009} for different aggregation windows of the same length, with the conclusion that the underlying process is stationary.
%
To quantify the convergence of the rescaled distributions to a single curve, we measure this convergence with the help of the $L^2$ distance between two distributions. The $L^2$ distance between two scalar functions $f$ and $g$ defined over the same domain $D \subseteq \mathrm{R}$ as
\begin{equation}
d(f,g) = \left(\int_{D} \left[f(x)-g(x)\right]^{2}dx\right)^{\frac{1}{2}},
\end{equation}
To measure the convergence, we have successively calculated the distance between the rescaled degree (weight) distributions of networks aggregated over an interval of length $t$ and networks aggregated over twice longer intervals of length $2t$. 
In Figure 8, it is seen that the $L^2$ distance between the degree distributions decreases and converges to a roughly constant value already for aggregation intervals longer than a few days. Hence, already after a short period, a sufficient amount of data has been collected to correctly estimate the stationary degree distribution of the network, and the shape of the degree remains similar for longer intervals. The weight distribution displays a slightly slower convergence, and is still slowly changing even for aggregation intervals of around 3 months; this is also related to the curvature seen in the evolution of the average link weight (Fig. 3 d).There are several possible explanations for this slower convergence: first, the number of links is increasing and each newly added link brings in a unit of weight. Second, the weights of existing links may display very different time dynamics, and third, the growth of link weights is affected by the burstiness of the call sequences and the resulting long inter-call times. Nevertheless,
even the weight distributions do not change much, and care is needed only when interpreting the weight distributions of call networks aggregated over very short periods of time.

\section*{On the effects of aggregation window placement}

In all analysis so far, we have assumed that the exact placement of the aggregation window, \emph{i.e.}~the time point of its beginning, plays no role in the results. However, as the characteristic daily and weekly patterns of Fig. 2 b) indicate, the overall level of call activity in the network displays large variations by hour and day, and this is expected to have an effect on the aggregated networks, at least on shorter time scales. In addition, there may be
less trivial effects if the actual behavioural patterns of individuals -- affecting \emph{who} they call -- are also time-dependent. In this final section, we will address these issues.

Let us first focus on short time scales, and illustrate the growth of the network with a 2D heat map plot, displaying the number of nodes in aggregated networks as a function of the aggregation time and the beginning point of the aggregation window (Fig. 9). The daily and weekly pattern is clearly visible in the plot -- the network grows fastest during weekdays, especially Fridays. The growth is slowest when the aggregation begins on Saturdays and especially Sundays; the difference between Saturdays and Sundays is fairly large. However, this observation might be explained by the network-wide variation of the call frequency alone.

\begin{figure}[t]
\begin{center}
\includegraphics[width=0.5\linewidth]{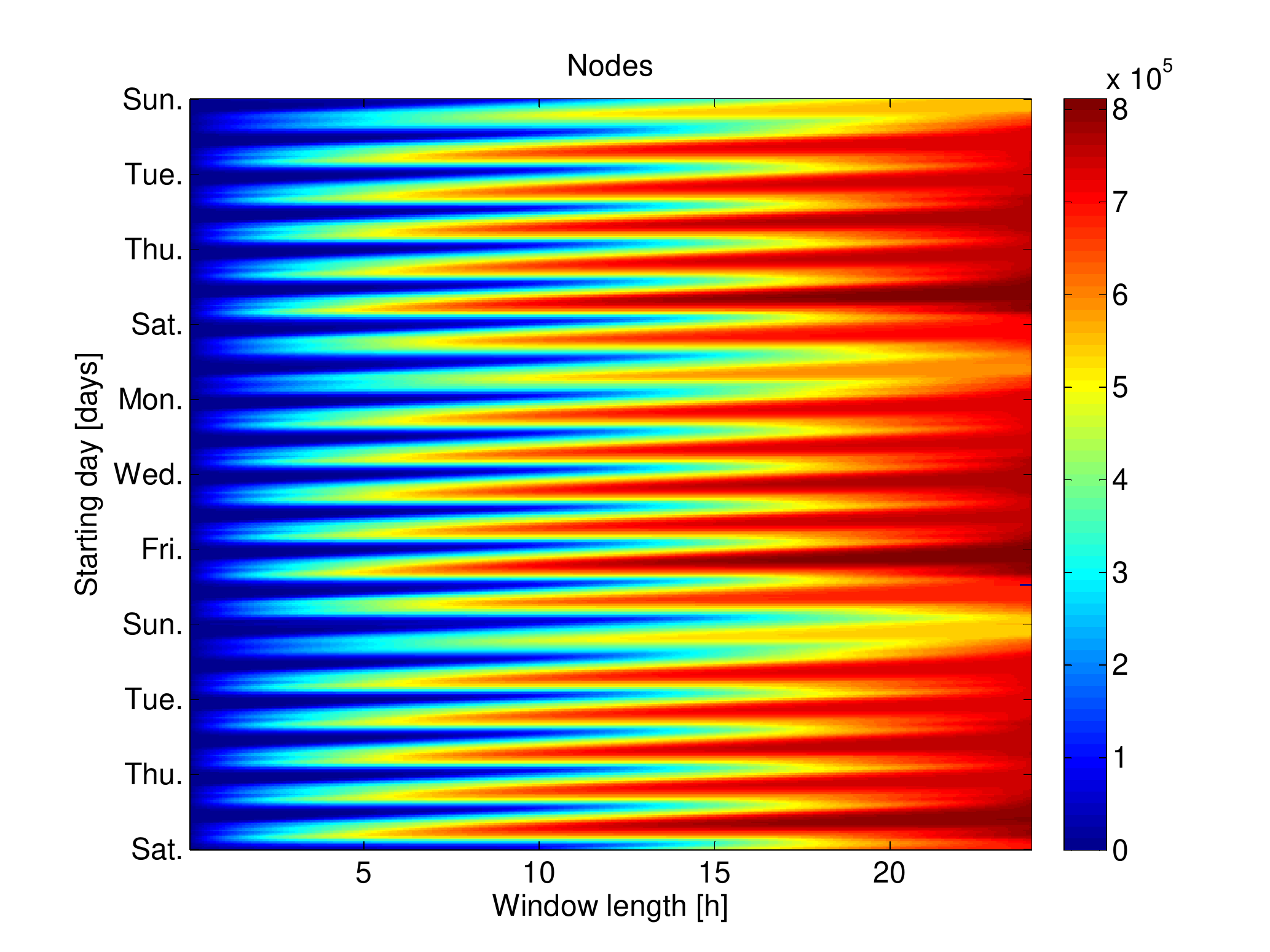}
\caption{The number of nodes in the aggregated networks as a function of the aggregation interval length (horizontal axis, in hours) and the beginning point of aggregation (vertical axis). The vertical axis runs from top to bottom, and the first time point is early Sunday, just after midnight.}
\label{fig:nodes3d}
\end{center}
\end{figure}

 In order to have a closer look at the actual call patterns of individuals, we monitor the growth of the giant connected component of the network. As seen in Fig. 5 b), the links that emerge early on in the aggregation process typically have high overlap, \emph{i.e.} are associated with dense network neighbourhoods; however, the overlap also displays a time-dependent pattern. Roughly speaking, if the majority of calls in a given window is directed towards high-overlap individuals that are part of the same neighbourhood or cluster, the giant component should grow more slowly than if the calls are directed towards far-away nodes. We measure this effect by monitoring the average size of the emerging largest connected component; for reference, we also calculate their size relative to those in the time-shuffled reference ensemble. In this ensemble, the exact times of all calls are randomly reshuffled, so that the links of the reference ensemble networks have the same number of calls as the original networks, but their timings are now uncorrelated, with the exception of the  the daily/weekly call frequency pattern at the network level. Figure 10 displays the absolute and relative size of the giant component similarly to the 2D heat map plot for nodes. The absolute size displays a clear daily and weekly pattern much akin to that seen for the number of nodes in Fig. 9. However, the behaviour of the average giant component size relative to the time-shuffled reference ensemble (lower panel) reveals an interesting feature: when the aggregation begins in the weekends, especially Sundays, the giant component grows more slowly than it does in the reference. Likewise, for weekdays, it grows far faster than it does in the reference. This points towards different behavioural modes -- weekday calls are frequently related to links joining nodes that would otherwise remain disconnected within the aggregation window, whereas weekend calls appear to relate to high-overlap links and dense clusters and thus contribute less to the growth of the largest connected components. In other words, friends and relatives with shared social circles are called more frequently in the weekends.

\begin{figure}[tb]

\begin{center}
\includegraphics[width=0.5\linewidth]{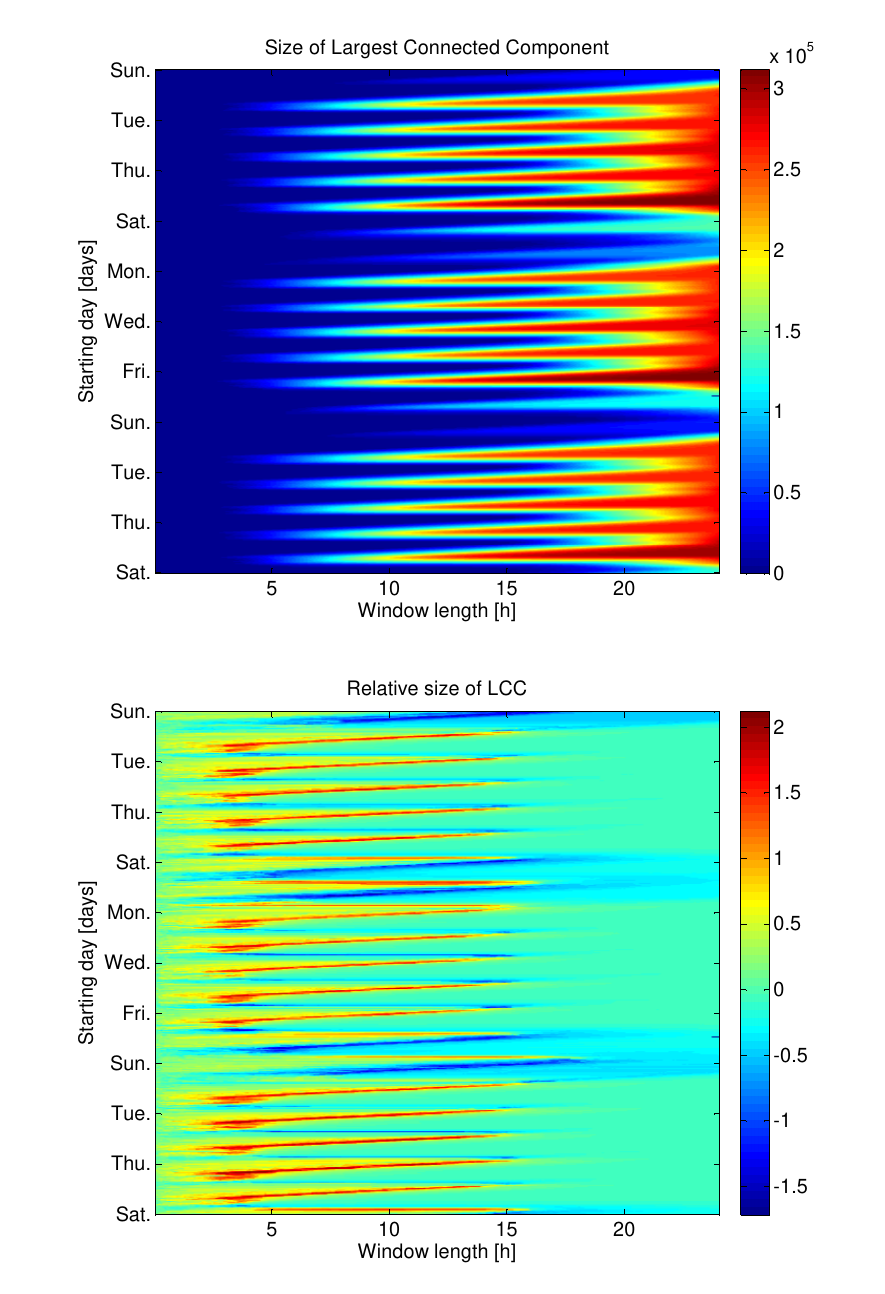}
\caption{The absolute (top) and relative (bottom) size of the largest connected component as a function of the aggregation time (horizontal axis) and the beginning point of aggregation (vertical axis). The relative values of the bottom panel are are the logarithm of the ratio between the observed largest connected component and the shuffled reference model, where call times do not correlate with network structure.}\label{fig:GCC}
\end{center}
\end{figure}

\begin{figure}[t!]
 \begin{center}
  \includegraphics[width=0.7\linewidth]{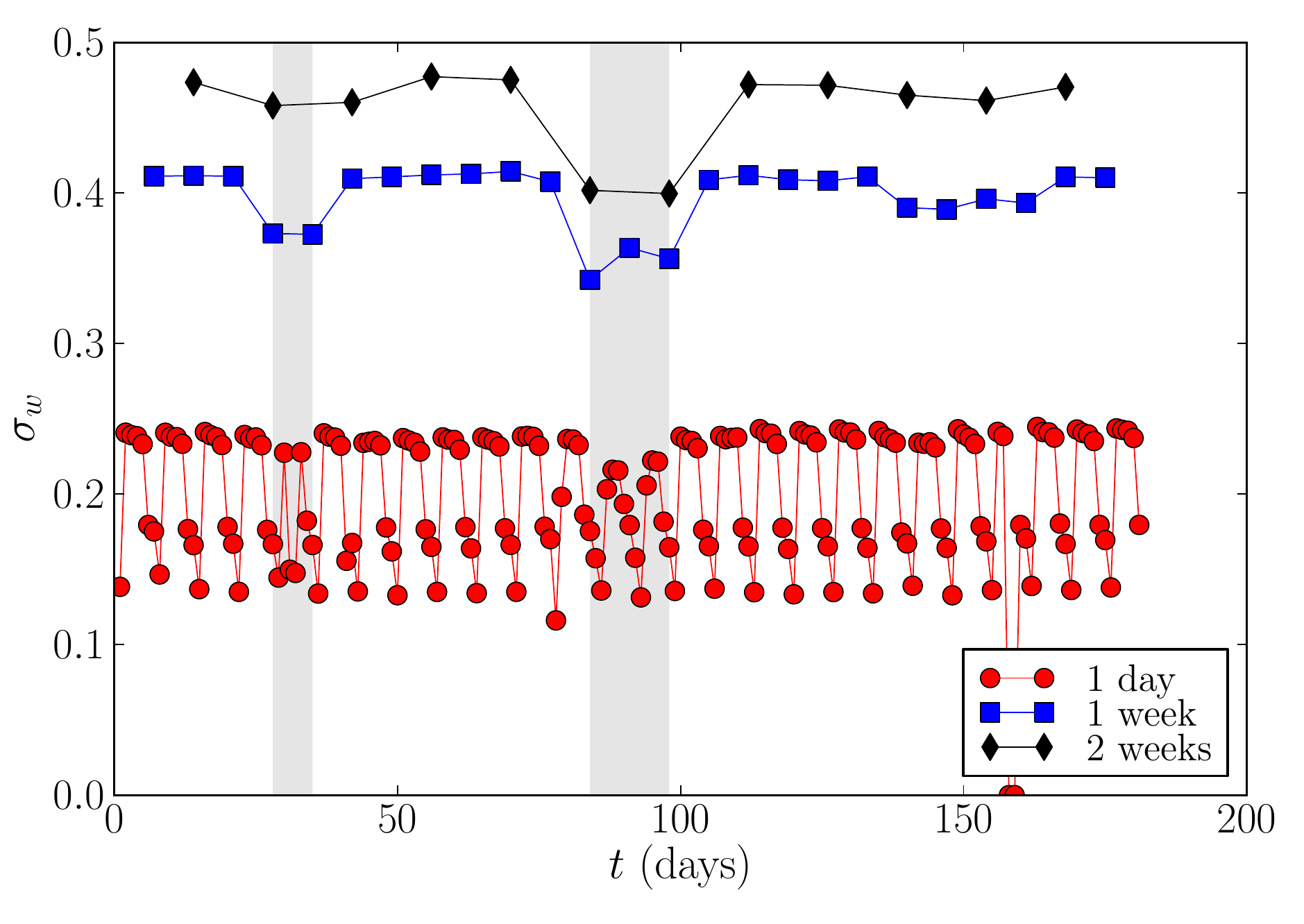}
 \end{center}
 \caption{Similarity between networks corresponding to two consecutive aggregation windows of 1 day, 1 week, and 2 weeks. The two shaded areas correspond to national holiday periods: the autumn holiday (left), and holidays around Christmas and New Year (right).}
 \label{fig:similarity_time}
\end{figure}

Changes in behaviour very similar to the daily and weekly patterns also appear on longer time scales, as human communication patterns are affected \emph{e.g.}~by holiday periods. In order to observe such behavioural changes in the aggregated networks, we divide the data into non-overlapping time windows of 1 day, 1 week, and 2 weeks. We then aggregate networks corresponding to each time window, and monitor the similarity of the networks in two consecutive windows as a function of time. For this, we employ a weighted generalization of the Jaccard index of Eq.~(\ref{eq:sim}):
\begin{equation}
\sigma_w=\frac{  \sum_{(i,j) \in E} \min(w_{ij}^{1},w_{ij}^{2})}{ \sum_{(i,j) \in
E} \max(w_{ij}^{1},w_{ij}^{2})},
\end{equation}

where $w_{ij}^1$ and $w_{ij}^2$ are the weights of links between nodes $i$ and $j$ in the consecutive windows ($w_{ij}=0$ if there is no link). 
Fig. 11 displays the weighted network similarities for the three different window sizes as a function of time. No long-term trends can be observed in the similarities that fluctuate around roughly constant values. The 1-day similarity displays a clear daily and weekly pattern, in line with our observations on the connected component sizes. Here, for each week, the consecutive windows corresponding to Monday to Thursday have highest similarities, while similarities are far smaller between the 1-day networks for Friday, Saturday and Sunday. Some larger changes can be seen already for the 1-day measure during the holiday periods (autumn holiday, Christmas season) indicated by the grey shaded areas. The differences corresponding to holiday periods become much more apparent when aggregation window sizes of 1 week and especially 2 weeks are used, as both measures display a clear drop, especially around the Christmas holiday season. Thus, the calling patterns of individuals are clearly different during such holiday seasons, and this is reflected in  the respective aggregated networks that are different from networks aggregated outside the holiday periods.

\section*{Conclusions}

In many cases, complex networks studied in the literature are constructed by aggregating links or sequences of interactions between the constituent nodes over some period of time, often limited by the availability of data, and their static structural features are then studied. The effects of the aggregation interval length and placement have been discussed only rarely (\cite{Holme2003,Gautreau2009}). In order to shed some insight into this problem, we have investigated the structural features of mobile telephone call networks aggregated over aggregation intervals of increasing length. To ensure that the results are not affected by churn, \emph{i.e.} customers leaving and subscribing to the operator, we only considered customers whose subscriptions did not change over the entire data interval from Oct 1st, 2006 to March 31th, 2007. 

Evidently, there several dynamical mechanisms and inhomogeneities that affect the features of networks aggregated over different time intervals, from broad distributions of numbers of calls on links to burstiness-related long inter-call times and dynamical changes in the network itself, and disentangling the effects of such features is not possible on the basis of time-stamped data alone. Thus the resulting networks display properties that arise from the interplay of such features associated with multiple time scales, and the question of a correct"or proper aggregation interval length is ill-posed. However, on the basis of our analysis, some statements about the general emergence of network features can be made. First, because of the broad link weight distribution and Granovetterian weight-topology correlations, where strong links are associated with dense neighbourhoods, the seeds of the underlying community structure are visible in aggregated networks already for rather short aggregation intervals of $\sim$1 week: the clustering coefficient of the network peaks at around 9 days, and the earlier a link is observed, the more likely it is to participate in a dense neighbourhood in the final network aggregated over the available data period, as seen by monitoring the neighbourhood overlap of the links.  However, at the same time, although the growth of the number of nodes saturates fairly early, the number of links and the average degree of nodes keep on growing even for long aggregation intervals. This suggests that for short windows, the cluster and community structures dominate, whereas for longer windows, the contribution of both "weak" links and links that are practically random, \emph{i.e.} arise from one-off-calls, increases. When networks from consecutive windows of different lengths are compared, they are seen to be maximally similar at a length of $\sim$30 days; this can be considered as the time scale of the recurrent, stable links, beyond which the weaker links start to have a considerable effect on network structure. The scaled degree and weight distributions become stationary already for short time intervals of a few days or weeks, respectively.

As the above results are from one dataset only, it is worth considering how general they are. As there are common underlying features of social networks -- broad tie strength distributions and the Granovetterian relationship between tie strengths and topology -- we believe that the fast emergence of clusters of strong links followed by increasing numbers of weaker links not associated with triangles is a general feature that holds across different communication networks. Likewise, one may assume that the time scale for obtaining stablest networks ($\sim 30$ days in our case) should remain roughly similar. However, in both cases, the exact numbers for the characteristic time scales might differ as they may also be affected by the overall call activity level. We also believe that the collapse of scaled distributions, indicating stationarity in the underlying processes, should be observable in other data sets too. 

In addition to the effects of the aggregation window length, we have shown that comparing networks aggregated over windows of different placement can yield insight into the dynamic features of the behavioural patterns of individuals. The differences in the growth of the largest connected component point towards different behavioural modes in the weekends and during weekdays, where weekend calls are more frequently related to high-overlap links and dense clusters, and thus build the largest connected component more slowly; weekday calls play the role of "topological shortcuts" in the aggregation process and more rapidly give rise to overall network connectivity. Additionally, we have observed very different calling patterns during holiday periods, giving rise to aggregated networks that significantly differ from the networks constructed from data outside the holiday periods. Thus, the aggregation interval placement matters, and care should be taken when interpreting the structural features of networks constructed from data that involves holidays or other special periods.

\bigskip

\section*{Author's contributions}
GK, MK, SB, VB and JS designed the research and analysis, GK prepared the data, GK and SB performed the analysis, GK, MK and JS wrote the paper

\section*{Acknowledgements}
  \ifthenelse{\boolean{publ}}{\small}{}
MK and JS acknowledge financial support of the Future and Emerging Technologies (FET) programme within the Seventh Framework Programme for Research of the European Commission, under FET-Open grant number: 238597 (project ICTeCollective). GK acknowledges support from the Concerted Research Action (ARC) ``Large Graphs and Networks'' from the ``Direction de la recherche scientifique - Communaut\'e fran\c{c}aise de Belgique''. The scientific responsibility rests with its authors.

\newpage
{\ifthenelse{\boolean{publ}}{\footnotesize}{\small}
 \bibliographystyle{bmc_article}  
  \bibliography{evnet} }


\ifthenelse{\boolean{publ}}{\end{multicols}}{}

\end{bmcformat}

\end{document}